\documentclass[showpacs,amssymb,aps,twocolumn,prb]{revtex4}

\usepackage{graphicx}
\usepackage{amsmath}

\newcommand{\beq}{\begin{equation}}
\newcommand{\eeq}{\end{equation}}
\newcommand{\bea}{\begin{eqnarray}}
\newcommand{\eea}{\end{eqnarray}}

\newcommand{\eps}{\epsilon}

\begin{document}

\title{Comment on: ``Zero temperature conductance of parallel T-shape double quantum dots'' [Physica E 39 (2007) 214, arXiv:0708.1842v1]}
\author{P. S. Cornaglia}
\affiliation{Instituto Balseiro and Centro At\'omico Bariloche, Comisi\'on
Nacional de Energ\'{\i}a At\'omica, 8400 San Carlos de Bariloche, Argentina.}

\date{\today}

\begin{abstract}
In a recent paper [Physica E {\bf 39} (2007) 214, arXiv:0708.1842v1] Crisan, Grosu 
and Tifrea revisited the problem of the conductance through a double 
quantum dot molecule connected to electrodes in a T-shape configuration.
The authors obtained an expression for the conductance that
disagrees with previous results in the literature. 
We point out an error in their derivation of the conductance formula
and show that it gives unphysical results even for non-interacting
quantum dots.
\end{abstract}

\pacs{72.15.Qm, 73.23.-b, 73.63.Kv}

\maketitle

The transport properties of double-quantum-dot (DQD) molecules 
connected to electrodes in a T-shape configuration
have been the subject of several investigations
using both numerical and analytical methods.
\cite{Kim2001,Cornaglia2005a,Zitko2006,Karrasch2006}
 In these devices an {\it active} dot is coupled to electrodes in a field-effect-transistor geometry and a 
second dot is {\it side coupled} to the first one through a hopping term. 

In this particular geometry, an exact expression for the zero-temperature 
conductance as a function of the charge in the dots can be obtained assuming a Fermi liquid ground 
state. It reads~\cite{Cornaglia2005a}
\beq \label{eq:cond}
G=\frac{G_0}{2}\sum_\sigma \sin^2[\pi(n_{A\sigma}+n_{S\sigma})],
\eeq
where $G_0=2e^2/h$ is the quantum of conductance and $n_{A\sigma}$
($n_{S\sigma}$) is the number of electrons with spin $\sigma$
in the active (side coupled) dot. 
This expression was obtained in Ref.~\onlinecite{Cornaglia2005a} 
using Meir and Wingreen\cite{Meir1992} formalism
and exact Fermi liquid relations in the wide-band limit.

Recently, Crisan, Grosu, and Tifrea (CGT)\cite{Crisan2007} revisited this problem
using the same methods and obtained a conductance formula that contains an additional term
in the r.h.s. of Eq.~(\ref{eq:cond}).
This additional term arises, however, due to an error in CGT's derivation.
In what follows we show that CGT used an incorrect form for the 
hybridization function that describes the coupling of the DQD to the electrodes.
As a consequence, their result is invalid and leads to unphysical results even for
non-interacting quantum dots.

The Hamiltonian of the DQD device is given by 
\begin{equation} \label{eq:hamilt}
  H=H_D+H_E+H_{DE}\;,
\end{equation}
where
\begin{eqnarray}
H_D &=& \sum_{\ell=A,S} \left[U_\ell n_{\ell \uparrow} n_{\ell
\downarrow}+ \varepsilon_\ell (n_{\ell \uparrow} +n_{\ell
\downarrow})\right] \nonumber\\ &-& t\sum_\sigma
\left(d^\dagger_{A\sigma} d_{S\sigma} +d^\dagger_{S\sigma}
d_{A\sigma}\right).
\end{eqnarray}
describes the $A-S$ quantum dot molecule and 
\begin{equation}
 H_{E}=\sum_{{\bf k},\sigma,\alpha} \varepsilon_{k\alpha}\;
c^{\dagger}_{{\bf k} \sigma \alpha} c^{}_{{\bf k} \sigma
\alpha}\;\left(\alpha={\rm L,R}\right)\;,\nonumber
\end{equation}
is the Hamiltonian of two non-interacting source and drain electrodes.
The coupling between the molecule and the electrodes is described by the last
 term in the Hamiltonian, 
\begin{equation}
H_{DE}=\sum_{k,\sigma,\alpha}\;V_{k\alpha,A}\left(d^{\dagger}_{A\sigma}
\;c_{k\sigma\alpha} + H.c.\right)\;. \nonumber
\end{equation}
From here on we set the Fermi level at zero and omit spin indices, for the sake of simplicity.
Meir and Wingreen \cite{Meir1992} derived a general formula for the current 
through an interacting region coupled to non-interacting leads\cite{Meir1992}
\begin{equation}\label{eq:current}
J=-\frac{2e}{h}\int d\epsilon [f_L(\epsilon
)-f_R(\epsilon)]\text{Im}[\text{Tr}\left\{ {\bf \Gamma}(\eps){\bf
G}^r(\eps)\right\}].
\end{equation}
where $[{\bf G}^r(t-t^\prime)]_{\ell,\ell^\prime}=-i\langle d_{\ell^\prime}^\dagger(t^\prime) d_{\ell}(t) \rangle$ is the retarded contour-ordered Green's function of the central region (in our case the DQD),
$f_{L}(\omega)$ and $f_{R}(\omega)$ are the Fermi-Dirac distributions of the leads, and  ${\bf
\Gamma}(\epsilon)\equiv{\bf \Gamma}^{L}(\epsilon){\bf \Gamma}^{R}(\epsilon)/[{\bf \Gamma}^{L}(\epsilon)+{\bf \Gamma}^{R}(\epsilon)]$ 
is given by the generalized linewidth functions 
\beq \label{eq:gamma}
[{\bf \Gamma}^{L(R)}(\epsilon)]_{\ell,\ell^\prime}= 2 \pi \rho_{L(R)}(\epsilon)
V_{L(R),\ell}(\epsilon)V^*_{L(R),\ell^\prime}(\epsilon).
\eeq
Here $\rho_{L(R)}(\epsilon)$ is the electronic density
of states of the left (right) electrode, and $V_{L(R),\ell}(\eps)$ equals
$V_{kL(R),\ell}$ for $\eps=\eps_k$. In deriving Eq.~(\ref{eq:current}) Meir and Wingreen assumed ${\bf \Gamma}^{L}(\epsilon)\propto {\bf \Gamma}^{R}(\epsilon)$.

For the DQD in the T-shape configuration we have
\beq \label{eq:gamma2}
{\bf \Gamma}(\epsilon) \equiv \begin{pmatrix} \Gamma_{A,A}(\epsilon) & \Gamma_{S,A}(\epsilon)  \\
\Gamma_{A,S}(\epsilon) & \Gamma_{S,S}(\epsilon) \end{pmatrix}=\begin{pmatrix} \frac{\Delta(\eps)}{2} & 0\\
0&0\end{pmatrix},
\eeq
where $\Delta(\epsilon) = (\Gamma^L_{A,A}(\epsilon) + \Gamma^R_{A,A}(\epsilon))/2$.
Replacing Eq.~(\ref{eq:gamma2}) in Eq.~(\ref{eq:current}) we obtain for the zero-temperature conductance
\beq
G=G_0\, \pi\, \Delta(0)\, \rho_{AA}(0),
\eeq
where $\rho_{AA}(\eps)=-\frac{1}{\pi}Im[G_{AA}^r(\eps)]$
is the spectral density of the active dot.

For non-interacting dots ($U=0$), the DQD Green's function is given by
\beq
[{\bf G}_0^r(\epsilon)]^{-1}=\begin{pmatrix} \epsilon- \epsilon_A+i \Delta(\epsilon) & t\\
t & \epsilon-\epsilon_S \end{pmatrix},
\eeq
and the zero-temperature conductance reads
\beq
G=\mathcal{C}\, G_0 \frac{\epsilon_S^2 \Delta^2}{(\eps_A \eps_S -t^2)^2
+\eps_S^2\Delta^2 }.
\eeq
where we have defined $\Delta\equiv\Delta(0)$ and $\mathcal{C}= 4 \Gamma^L_{AA}(0)\Gamma^R_{AA}(0)/[\Gamma_{AA}^L(0)+\Gamma_{AA}^R(0)]^2$, which is equal to one for $\Gamma_{AA}^L(0)=\Gamma_{AA}^R(0)$. This expression can also be obtained using Landauer formalism. 

In their derivation, however, CGT used an expression for ${\bf \Gamma}(\eps)$ 
which is inconsistent with its definition [see Eq.~(\ref{eq:gamma})]
\beq\label{eq:gammaCGT}
{\bf \Gamma}_{CGT}(0)=\begin{pmatrix} -\Delta & it\\ it&0\end{pmatrix}. 
\eeq
Furthermore this expression is not hermitian as it should be, and has three matrix elements out of four different from zero which doesn't seem to correspond to any physical model.
Using Eq.~(\ref{eq:gammaCGT}) the conductance has the following form\endnote{Note that Eq.~(\ref{CGTformula}) is different from Eq.~(14) in
Ref.~\onlinecite{Crisan2007}, most probably due to some typos.}

\beq \label{CGTformula}
G_{CGT}=G_0\frac{\eps_S^2\Delta^2- 2 t^2\eps_A \eps_S +2
t^4}{(\eps_A \eps_S-t^2)^2
+\eps_S^2\Delta^2}.
\eeq
This formula has some clear inconsistencies and gives unphysical results. For $\Delta=0$ the dots are
decoupled from the electrodes and we should have $G=0$. Equation
(\ref{CGTformula}) gives, however,
a non-zero result that can be larger than $G_0$ or even negative  
\begin{equation}
G_{CGT}(\Delta=0)=\frac{2 t^2 G_0} {(t^2-\eps_A\eps_S)}\,.
\end{equation}
Furthermore, for $\Delta$ non-zero a negative
conductance ($\eps_A=-4, \eps_S=-4, \Delta=1, t=2$)
or a conductance larger than $G_0$ ($\eps_A=-4, \eps_S=-4, \Delta=1, t=6$)
is obtained. 
The additional term to Eq.~(\ref{eq:cond}) obtained by CGT is a consequence
of the use of this incorrect expression for the conductance and is therefore
spurious.

In summary, we have presented a derivation of the zero-temperature conductance
formula for a DQD device in a T-shape configuration and we have shown
that CGT's conductance formula is incorrect and leads to unphysical results.

We thank C.A. Balseiro for useful remarks and M. Crisan useful correspondence.

\bibliography{refs}

\end{document}